\def\includeappendix{yes}
\documentclass[
 reprint,
 superscriptaddress,
 amsmath,amssymb,
 aps,
 prl,
 showkeys
]{revtex4-1}
\usepackage{afterpage}
\usepackage{graphicx}
\usepackage{xcolor}
\usepackage{dcolumn}
\usepackage{bm}
\usepackage{hyperref}
\usepackage{cancel}
\usepackage{etoolbox}
\usepackage{upgreek}
\newcommand\rr{\mathbf{r}}
\newcommand\EE{\mathbf{E}}
\newcommand\HH{\mathbf{H}}
\newcommand\DD{\mathbf{D}}
\newcommand\BB{\mathbf{B}}
\newcommand\rte{r_{\text{TE}}}
\newcommand\rtm{r_{\text{TM}}}
\newcommand\hcons{h_{\text{c}}}
\newcommand\hflip{h_{\text{f}}}

\newcommand{\upsub}[1]{\sb{\mathrm{#1}}}
\newcommand{\upsup}[1]{\sp{\mathrm{#1}}}

\begingroup\lccode`~=`_\lowercase{\endgroup\let~\upsub}
\begingroup\lccode`~=`^\lowercase{\endgroup\let~\upsup}

\AtBeginDocument{
  \catcode`_=12 \catcode`^=12
  \mathcode`_="8000
  \mathcode`^="8000
}

 \ifdefstring{\includeappendix}{no}{\usepackage{xr}\externaldocument{main_appendices_cavity_double_array_cd}}{}

\begin{document}

\title{Helicity-preserving optical cavity modes for enhanced sensing of chiral molecules}

\author{Joshua Feis}
\affiliation{Institute of Applied Physics, Karlsruhe Institute of Technology, 76128 Karlsruhe, Germany}
\author{Dominik Beutel}
\affiliation{Institute of Theoretical Solid State Physics, Karlsruhe Institute of Technology, 76128 Karlsruhe, Germany}
\author{Julian K\"opfler}
\affiliation{Institute of Applied Physics, Karlsruhe Institute of Technology, 76128 Karlsruhe, Germany}
\affiliation{Institute of Nanotechnology, Karlsruhe Institute of Technology, 76021 Karlsruhe, Germany}
\author{Xavier Garcia-Santiago}
\affiliation{Institute of Nanotechnology, Karlsruhe Institute of Technology, 76021 Karlsruhe, Germany}
\affiliation{JCMWave GmbH, 14050 Berlin, Germany}
\author{Carsten Rockstuhl}
\affiliation{Institute of Theoretical Solid State Physics, Karlsruhe Institute of Technology, 76128 Karlsruhe, Germany}
\affiliation{Institute of Nanotechnology, Karlsruhe Institute of Technology, 76021 Karlsruhe, Germany}
\author{Martin Wegener}
\affiliation{Institute of Applied Physics, Karlsruhe Institute of Technology, 76128 Karlsruhe, Germany}
\affiliation{Institute of Nanotechnology, Karlsruhe Institute of Technology, 76021 Karlsruhe, Germany}
\author{Ivan Fernandez-Corbaton}
\affiliation{Institute of Nanotechnology, Karlsruhe Institute of Technology, 76021 Karlsruhe, Germany}

\date{\today}

\begin{abstract}
	Researchers routinely sense molecules by their infrared vibrational ``fingerprint'' absorption resonances. In addition, the dominant handedness of chiral molecules can be detected by circular dichroism (CD), the normalized difference between their optical response to incident left- and right- handed circularly polarized light. Here, we introduce a cavity composed of two parallel arrays of helicity-preserving silicon disks that allows one to enhance the CD signal by more than two orders of magnitude for a given molecule concentration and given thickness of the cell containing the molecules. The underlying principle is first-order diffraction into helicity-preserving modes with large transverse momentum and long lifetimes. In sharp contrast, in a conventional Fabry-Perot cavity, each reflection flips the handedness of light, leading to large intensity enhancements inside the cavity, yet to smaller CD signals than without the cavity.
\end{abstract}

\maketitle
The interplay of molecules with nearby metallic or dielectric nanostructures allows for effectively enhancing light-molecule interaction. Examples are surface-enhanced Raman scattering \cite{Jeanmaire1977,Kneipp1997,Stockman2006}, molecule infrared vibrational absorption resonances enhanced by plasmonic antennas \cite{Adato2009}, molecules in the near field of whispering-gallery-mode optical resonators \cite{Armani2007}, and largely enhanced molecular spontaneous emission rates in the vicinity of silver nanocubes \cite{Akselrod2014}. Furthermore, frequency-selective optical metasurfaces can effectively integrate an otherwise necessarily external optical spectrometer \cite{Tittl2018}.

However, all of the mentioned techniques are typically not able to sense the dominant handedness of chiral molecules. This aspect is important, though. For example, the majority of today’s drugs use molecules of only one handedness (i.e., one enantiomer). Sometimes, the wrong molecule handedness has perturbing side effects. Therefore, enhancing the CD signal is highly desirable, especially if only small quantities of analyte with small concentrations of molecules are accessible. Early on, it was expected that the interaction of chiral nanostructures with chiral molecules would be helpful \cite[\protect{Chap. 5}]{Schaeferling2016}\cite{Hentschel2012,Valev2013}. However, more recent work shows that the chirality of the nanostructure itself is rather unwanted \cite{Wu2014,Nesterov2016,Mohammadi2018,Graf2019}, since it distorts the CD measurements, for example, by producing a nonzero signal even for racemic solutions or achiral molecules. The work in Ref. \onlinecite{Graf2019} also shows that the achiral nanostructure should be helicity preserving, such that the near field and the scattered far field have the same handedness as the incident light. Along these lines, systems for enhancing CD signals have been reported \cite{Yoo2015,Mohammadi2018,Garcia-Guirado2018,Vazquez-Guardado2018,Solomon2019,Graf2019}. However, the spatial region of enhancement was restricted to only parts of the optical near field which extended over a few hundreds of nanometers or less. Therefore, merely a small fraction of molecules in a microfluidic channel would experience this enhancement, hence reducing the total CD enhancement factor to near unity, that is, to no beneficial effect at all. This aspect severely limits the practicability of these previous approaches. 

In this Letter, we design, describe, and characterize a cavity composed of two parallel dielectric metasurfaces. The cavity features approximately helicity-preserving modes with long lifetimes, leading to resonant CD enhancement factors exceeding one hundred, for cavity lengths of a few tens of micrometers. To achieve feedback, this cavity does not exploit the normal-incidence reflection off the metasurfaces. For an ordinary mirror, this reflection would convert left-handed into right-handed circularly polarized light and {\em vice versa}, destroying any positive effect of the cavity on the CD. The cavity rather exploits modes whose lifetime and helicity preservation are inherently large because their momentum is nearly parallel to the disk array plane. The long interaction times allow to selectively and resonantly enhance certain chiral molecular spectral resonances. 

We proceed by showing that planar cavity modes with large transverse momentum feature long lifetimes and helicity preservation. The latter is a generic property that also occurs for light reflected at grazing incidence off a dielectric half-space. We then design a tunable cavity using two diffracting silicon disk arrays and numerically analyze its CD enhancement performance. Finally, we show that the spatial regions of high enhancement inside the cavity are away from the near-field regions attached to the disks. To the best of our knowledge, this kind of cavity design has not been reported before.

A set of design requirements that lead to structures suitable for molecular CD enhancement has recently been formalized \cite{Graf2019}: Achirality, helicity preservation, and strong light-matter interaction. The paradigmatic system for enhancing light-matter interaction is a cavity. An achiral cavity with helicity-preserving modes with long lifetimes would be a distinctly advantageous system for CD enhancement. Helicity-preserving resonances in achiral systems are a challenging requirement, though. Such resonances are, in principle, perfect mixes of the two helicities. The eigenmodes of systems with spatial inversion symmetries will also be eigenmodes of at least one of those inversion operations, and can hence be classified by their inversion eigenvalue $\tau=\pm1$ \cite[p. 181]{Wigner1959}. For example, the electric and magnetic resonant modes in spheres are eigenstates of the parity operator, and the modes in planar waveguides are classified as TE or TM depending on their eigenvalue upon transformation by a mirror symmetry of the system. While the relevant inversion operator can be different in each system, the modes of achiral systems can always be written as the sum or subtraction of equal amplitude modes of well-defined helicity $\chi=\pm1$ \cite[Sec. 2.4.2]{FerCorTHESIS}: 
\begin{equation}
	\label{eq:heltau}
\sqrt{2}|\eta\ \tau\rangle = |\eta \ \chi=+1\rangle+\tau|\eta \ \chi=-1\rangle,
\end{equation}
where $\eta$ is a composite index that completes the characterization of the mode\footnote{For example, for spherical waves $\eta$ is composed by the frequency, total angular momentum, and angular momentum along a chosen axis.}. Helicity preservation is hence not expected from isolated eigenmodes of systems with spatial inversion symmetry. However, helicity-preserving modes will be achieved when a pair of $|\eta \ \tau=\pm 1\rangle$ modes becomes degenerate \cite[p. 4]{FerCor2012p}. Upon degeneracy, any linear combination of the $\tau=\pm1$ modes is also a modal solution, and the resonance preserves the polarization state of the excitation. This statement applies in particular to the pure helicity $\chi=\pm1$ combinations:
\begin{equation}
	\label{eq:tauhel}
\sqrt{2}|\eta\ \chi\rangle = |\eta \ \tau=+1\rangle+\chi|\eta \ \tau=-1\rangle.
\end{equation}
We now show that modes with large transverse momentum in planar cavities become helicity-preserving resonances. We start by arguing that degeneracy-induced helicity preservation is achieved quite generally in the reflection off planar systems in the limit of grazing incidence angle\footnote{\protect{General degeneracy, independent of anything else, is achieved if the system has electromagnetic duality symmetry. This is easily seen from the action of the generator of duality, the helicity operator $\Lambda$, on the $\tau$ eigenstates: $\Lambda\sqrt{2}|\eta \ \tau\rangle =\sqrt{2}\Lambda \left(|\eta \ +\rangle +\tau|\eta \ -\rangle\right)=\sqrt{2}\left(|\eta\ +\rangle-\tau|\eta\ -\rangle\right)=\sqrt{2}|\eta\ -\tau\rangle$. Since the $\tau=\pm$1 modes transform into each other by the action of the generator of a symmetry of the system, they must be degenerate. Systems whose materials have all the same impedance, $Z_i=\sqrt{\mu_i/\epsilon_i}=Z$ for all $i$, are dual symmetric \cite{FerCor2012p}. Dual systems are challenging to obtain, specially for near infrared and higher frequencies. As seen in Eq.~(\ref{eq:tetm}), $\rte=\rtm$ is achieved for all $\beta$ if and only if $Z_1=Z_2$ \cite{FerCor2012p}.}}. Let us consider a unit amplitude circularly polarized plane wave propagating in medium 1 and reflecting off a planar interface with medium 2. Both media are isotropic and homogeneous. We denote by $\beta$ the angle of incidence measured with respect to the normal. It follows from Eqs.~(\ref{eq:heltau})-(\ref{eq:tauhel}) that the reflected plane wave contains the two helicity components with coefficients
\begin{equation}
	\label{eq:hchf}
	\hcons=\frac{1}{2}\left(\rte+\rtm\right),\ \hflip=\frac{1}{2}\left(\rte-\rtm\right),
\end{equation}
for the preserved and flipped helicity component, respectively, where $\rte(\rtm)$ is the Fresnel reflection coefficient for the TE(TM) polarization, which depends on $\beta$ and on the material parameters. It is readily seen from the expressions of the Fresnel reflection coefficients that, when $\beta\rightarrow \pi/2$, then $\rte\rightarrow -1$ and $\rtm\rightarrow -1$, independently of the material parameters. See \ifdefstring{\includeappendix}{no}{Sec.~\ref{app:fresnel} of the Supplemental Material}{}\ifdefstring{\includeappendix}{yes}{Appendix \ref{app:fresnel}}{} for a short derivation of this known phenomenon. Importantly, the same limiting behavior, $\lim_{\beta\rightarrow \pi/2}\rte(\rtm)\rightarrow -1$, occurs in more general kinds of interfaces featuring (discrete) transverse translational symmetry, including anisotropic media and (diffracting) gratings \cite[Chap. 2.3]{Lekner2016}\cite{Nakayawa2011}. We therefore see from Eq.~(\ref{eq:hchf}) that the intensity of the flipped helicity component ($|\hflip|^2$) tends to zero as $\beta\rightarrow \pi/2$. This mechanism of helicity preservation as $\beta\rightarrow \pi/2$ can be seen on the one hand as the opposite of the complete helicity flipping for reflection when $\beta=0$, and on the other hand as the reflection counterpart of the perfect helicity preservation for transmission at $\beta=0$.

Let us now consider the modes of a slab waveguide. The effect discussed above suggests that, as $\beta\rightarrow \pi/2$, the TE and TM modes of the waveguide degenerate into helicity-preserving modes. That this is indeed the case can readily be seen from the expressions for the modal condition in a slab waveguide \cite[Eqs. 8.121-8.122]{Jackson1998}, which, in our notation, read as:
\begin{equation}
	\label{eq:slab}
	\begin{split}
		&2Lk\cos\beta +2\Phi_{\text{TE}/\text{TM}}= 2\pi l_{\text{TE}/\text{TM}},\\
		\Phi_{\text{TE}}&=-2\arctan\left(\sqrt{\frac{2\Sigma}{\cos^2\beta}-1}\right),\\
		\Phi_{\text{TM}}&=-2\arctan\left(\frac{1}{1-2\Sigma}\sqrt{\frac{2\Sigma}{\cos^2\beta}-1}\right), 
	\end{split}
\end{equation}
where $L$ is the thickness of the slab, $k$ is the wavenumber inside the slab, $l_{\text{TE}}(l_{\text{TM}})$ is an integer and $2\Sigma=1-\left(n_2/n_1\right)^2$, where $n_1(n_2)$ is the refractive index of the slab(exterior region). As before, when $\beta\rightarrow \pi/2$, then $\Phi_{\text{TE}}\rightarrow \pi$ and $\Phi_{\text{TM}}\rightarrow \pi$, and the TE and TM modes become degenerate. Therefore, modes of large transverse momentum become helicity-preserving resonances. This physical insight can be used for designing a system for enhanced CD measurements. 

\begin{figure}[h!]
		\includegraphics[width=\linewidth]{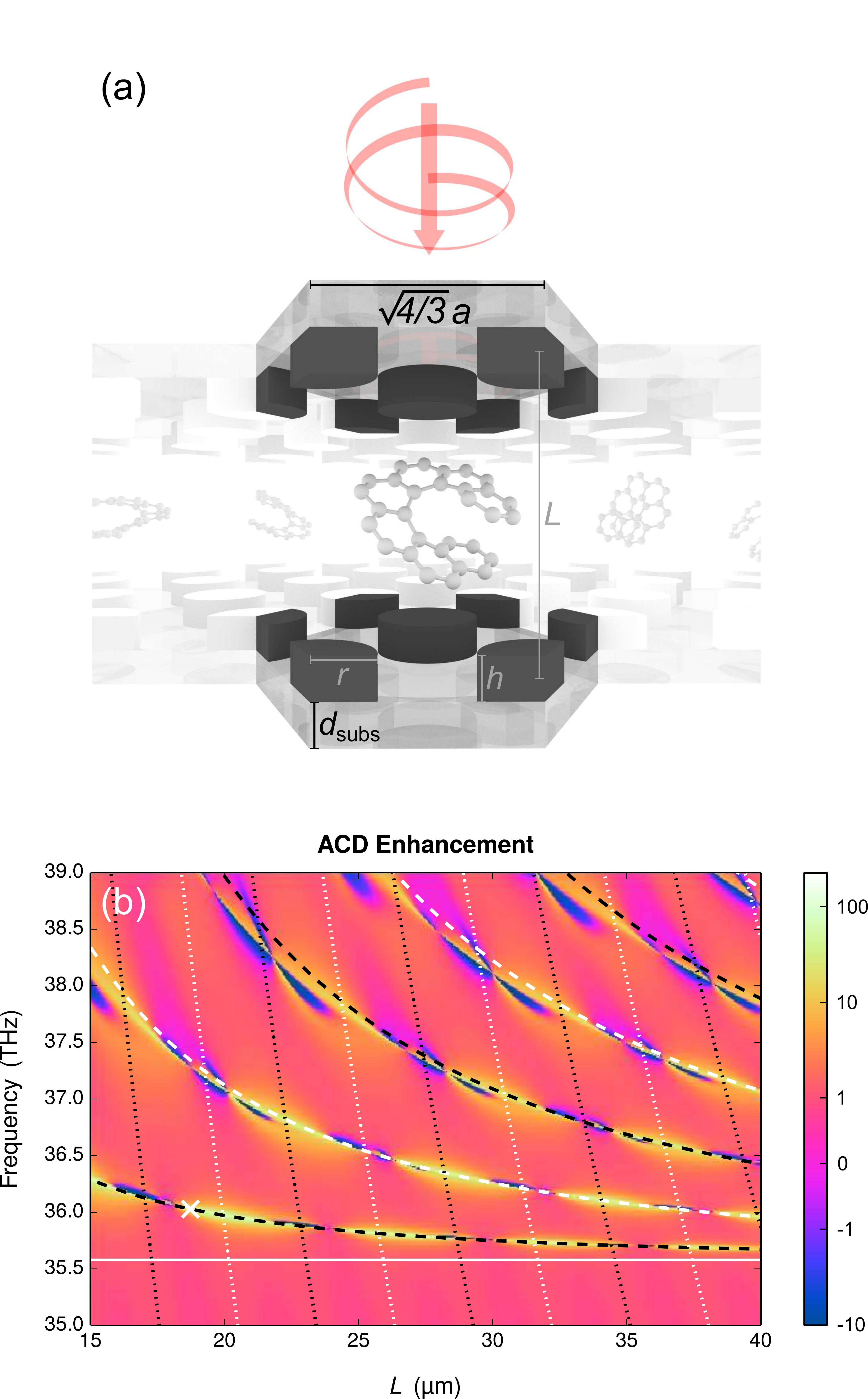}
		\caption{(a) Sketch of the cavity and illumination (not to scale). The dimensions are, in micrometers, $a=5.76$, $h=1.32$, $r=1.92$, and $d_{\text{subs}}=2.0$. The disks are made out of silicon and the dielectric substrates have $\epsilon_{\text{subs}}=2.14$. (b) Absorbance CD enhancement achieved by the system under perpendicular illumination as a function of frequency and cavity length ($L$). The scale is a symmetric logarithmic scale (see the text). The dashed (dotted) lines are approximated modal conditions for modes composed of the first (zeroth) array diffraction orders. The solid horizontal line marks the onset of diffraction in an $\epsilon_{\mathrm{r}}=2.14$ medium.\label{fig:lines} }
\end{figure}

Let us imagine that we place two planar substrates facing each other and leave a gap between them forming a cavity, which plays the role of the slab. The gap is to be filled with the solution of chiral molecules. We then need a means for exciting modes with large $\beta$, and then measuring the outgoing power. This is not trivial because large $\beta$ implies a large momentum component in a direction transverse to the cavity walls which, together with the refractive index of the solvent, prevent such modes to couple directly to electromagnetic waves propagating in air, whose momentum is insufficient. One possible solution is to add diffracting arrays for providing the necessary momentum matching for both excitation and measurement along the normal $\pm z$ directions. The structure that we propose is a cavity formed by two identical planar arrays of cylinders placed opposite each other at a distance $L$. The cylinders are arranged in the form of hexagonal lattices and are supported by finite-thickness substrates that seal the chamber. Figure~\ref{fig:lines}(a) shows a sketch of the cavity. The intended functionality is that, upon circularly polarized perpendicular illumination from the top array, a diffraction order excites an approximately helicity-preserving resonant mode in the cavity. The light then interacts with the molecules for some time determined by the lifetime of the resonance. Finally, again thanks to diffraction off the arrays, the light can couple back out of the cavity along the $\pm z$ directions, where detectors are placed.

While the modes in a cavity like the one that we propose are more complex than in a slab waveguide, they can still be classified as TE or TM. We can hence expect that, because the specular reflection from a single array behaves similarly to the case of a planar interface, the cavity modes with large transverse momentum will become helicity-preserving resonances. In \ifdefstring{\includeappendix}{no}{the Supplemental Material}{}\ifdefstring{\includeappendix}{yes}{Appendix \ref{app:figures}}{} we analyze the zeroth-order reflection coefficient for a single diffracting array of silicon disks embedded in an $\epsilon_{\text{r}}=2.14$ medium as a function of frequency, and incidence angle $\beta$. Figure~\ref{fig:reflectionsinglearray}(a) shows the helicity change measure $|\hflip|^2/\left(|\hcons|^2+|\hflip|^2\right)$ for the zeroth-order (specular) reflection. Figure~\ref{fig:reflectionsinglearray}(b) shows the real part of $\hcons$. The results meet the expectation: As $\beta\rightarrow \pi/2$, $\hcons\rightarrow -1$ and $|\hflip|^2\rightarrow 0$. Two distinct lines where the convergence is much slower can be observed. They correspond to modes of the single array. Importantly, we note that a mode with $\beta\rightarrow \pi/2$ automatically has a long lifetime because when $|\hcons|\rightarrow 1$, the coupling to the channels through which the energy can leave the mode tends to zero due to energy conservation. This implication is quantitatively verified in Fig.~\ref{fig:reflectionsinglearray}(c), which shows that the average number of internal reflections before the energy leaves the mode grows as $\beta\rightarrow \pi/2$. That this happens in a diffracting array is quite remarkable. 

We will now investigate the CD enhancement performance of the design. We will use the following definition of Absorbance CD (ACD)
\begin{equation}
\label{eq:relcd}
    \mathrm{ACD} = \frac{P_+ - P_-}{2P_0},
\end{equation}
where $P_+$ and $P_-$ are the outgoing powers for left-handed (``+'') and right-handed (``-'') circularly polarized incident light, for identical incident powers equal to $P_0$. The outgoing power includes both transmitted and reflected fields. The ACD measures the differential absorption of the molecules relative to the incident power. We also consider the Transmittance CD (TCD), defined as $\left(P^{\text{fwd}}_{+}-P^{\text{fwd}}_{-}\right)/\left(P^{\text{fwd}}_{+}+P^{\text{fwd}}_{-}\right)$, where only the transmitted powers $P^{\text{fwd}}_{\pm}$ are detected. The ACD(TCD) enhancement factor is defined by the ACD(TCD) signal with the arrays divided by that without the arrays.

The chiral solution in the cavity is modeled by the constitutive relations
\begin{equation}
	\label{eq:consrel}
	\DD = \epsilon_0 \epsilon_{\mathrm{r}} \EE + \mathrm{i}\frac{\kappa}{c_0} \HH \ \text{and} \  \BB = \mu_0 \mu_{\mathrm{r}} \HH - \mathrm{i}\frac{\kappa}{c_0} \EE, 
\end{equation}
where $\epsilon_{\mathrm{r}} = 2.14 + \mathrm{i}10^{-4}$,  $ \mu_{r} = 1 + \mathrm{i}10^{-4}$, and $\kappa = \mathrm{i} 10^{-4}$. The dielectric permittivity of the substrates is assumed to be equal to $\epsilon_{\text{subs}}=2.14$. These material parameters are realistic choices for substrates and many solvents commonly used in infrared measurements of chiral molecules \cite{Yadav2013}. The cylinders are made out of silicon with $\epsilon_{\text{cyl}} = 11.9$ and $\mu_{\text{cyl}} = 1$. 
Regarding the numerical analysis, we refer the reader to \ifdefstring{\includeappendix}{no}{Sec.~\ref{app:code} of the Supplemental Material}{}\ifdefstring{\includeappendix}{yes}{Appendix \ref{app:code}}{} for details. Figure~\ref{fig:lines}(b) shows the ACD enhancement in a symmetric logarithmic scale\footnote{The logarithm of the quantity for positive values larger than 1, the negative logarithm of the absolute value of the quantity for negative values smaller than -1 and a linear interval from -1 to 1.} as a function of the array-to-array distance $L$ and the illumination frequency $f$. The ACD enhancement is larger than thirtyfold in several areas along a set of curved lines. The enhancement is higher around the lowest lying line where it reaches a maximum value of $126$ at $(L,f)=(18.73\ \upmu\text{m},36.03\ \text{THz})$. This point is marked with a white cross. The onset of diffraction, where $\beta=\pi/2$, occurs at $f=35.58$ THz and is marked by the white continuous line. The curved dashed lines are computed using 
\begin{equation}
	\label{eq:lines_simple}
		L\sqrt{1-\left(\frac{\lambda}{a}\right)^2}=\frac{\lambda}{2}(l_1-1),
\end{equation}
for integer $l_1-1\in\{1,..,6\}$, where $\lambda$ is the wavelength in an $\epsilon_{\text{r}}=2.14$ medium, and $\frac{\sqrt{4}{3}}a$ is the center-to-center distance between the disks. Lower values of $l_1-1$ correspond to lines closer to the diffraction limit. We show in \ifdefstring{\includeappendix}{no}{Sec.~\ref{app:array} of the Supplemental Material}{}\ifdefstring{\includeappendix}{yes}{Appendix \ref{app:array}}{} that Eq.~(\ref{eq:lines_simple}) is the approximated modal condition for the first six frequency degenerate diffraction orders from the hexagonal lattice upon normal incidence. The $-1$ on the right-hand side of Eq.~(\ref{eq:lines_simple}) arises from the phase of $\pi$ acquired upon reflection at large incidence angles. We can also observe in Fig.~\ref{fig:lines}(b) that the enhancement has some dips along the lines. These are due to coupling between the first diffraction orders and the zeroth diffraction orders, whose modal equation is [see Eq.~(\ref{eq:zero})]:
\begin{equation}
	\label{eq:lines_simple_zero}
		L=\frac{\lambda}{2}l_0.
\end{equation}
The dotted lines in Fig.~\ref{fig:lines}(b) correspond to Eq. (\ref{eq:lines_simple_zero}) for integer $l_0\in\{6,\ldots,15\}$. The enhancement dips do not occur at every crossing of the lines determined by Eq. (\ref{eq:lines_simple}) and Eq. (\ref{eq:lines_simple_zero}), but rather in an alternating fashion. As shown in \cite[Sec. 2.3c]{Chang-Hasnain2012}, two different kinds of modes can couple when the phase that they accumulate in a half round-trip is an integer multiple of $2\pi$. We show in Sec.~\ref{app:array} that they can couple when $l_1-l_0$ is even or, equivalently, when $(l_1-1)-l_0$ is odd. In Fig.~\ref{fig:lines}(b), the crossings of dashed and dotted lines of the same color mark points where the two kind of modes can couple. At these points, the grazing incidence angle properties of the first-order modes cease to exist in the resulting mixed modes, and the enhancement disappears. The first-order modes are unperturbed in the crossings of dashed and dotted lines of different color. The maximum enhancement occurs at $L=18.73$~$\upmu$m in the lowest lying line\footnote{The increasingly finer enhancement linewidths as $\beta\rightarrow \pi/2$ makes it increasingly more difficult to reliably evaluate the enhancements for angles very close to 90 degrees.}, which corresponds to $\beta\approx 81$ degrees [Eq.~(\ref{eq:beta})].  

It should be noted that the first six diffraction orders can propagate in the cavity and in the substrates, but become evanescent in vacuum\footnote{In vacuum, the diffraction-free region extends up to $\approx 52$ THz.}. The system is hence nondiffracting for external illumination and measurement: All the information about the internal light-matter interaction available outside the system is contained in the zeroth transmission and reflection orders. 

In many CD measurement setups, only the transmitted power is detected. Figure~\ref{fig:cdehnforward} in \ifdefstring{\includeappendix}{no}{the Supplemental Material}{}\ifdefstring{\includeappendix}{yes}{Appendix \ref{app:forwardonly}}{} shows the Transmittance CD enhancement, where features and enhancement values very similar to those seen in Fig.~\ref{fig:lines}(b) can be observed. While in a few regions the TCD enhancement can be partly due to a sharp decrease in transmittance, Fig.~\ref{fig:t0} shows that this is not the general case, and that the averaged forward transmitted power $\left(P^{\text{fwd}}_++P^{\text{fwd}}_-\right)/2$ is relatively large for many high enhancement regions. We also note that the incoupling causes some helicity nonpreservation\footnote{Our numerical calculations show that the transmission from perpendicular incidence to the first diffraction order features a helicity preservation degree of $\approx 82\%$ between 35.6 and 38.5~THz.} 

\begin{figure}[h!]
		\includegraphics[width=0.8\linewidth]{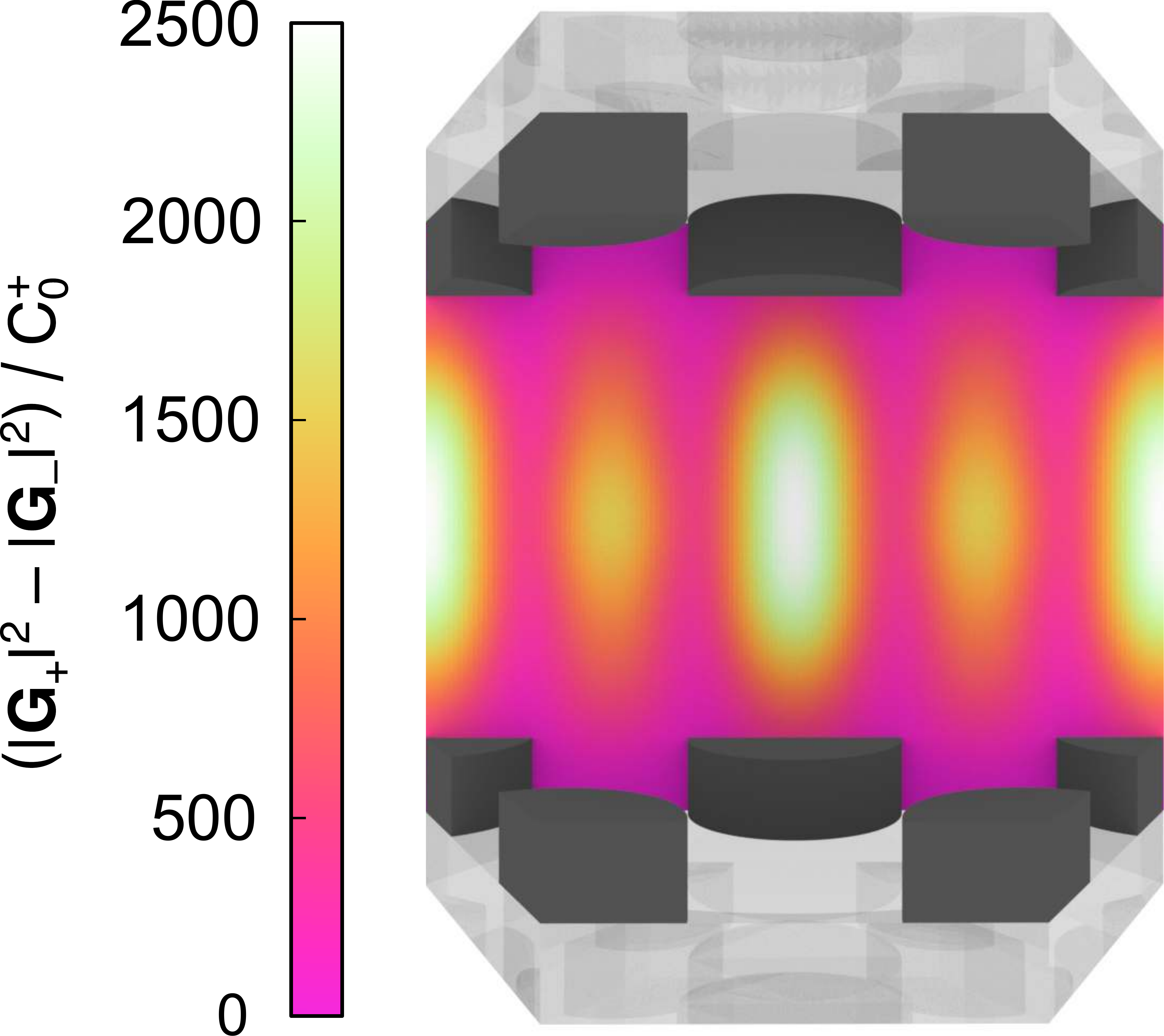}
		\caption{Spatial intensity maps of the helicity density enhancement inside the cavity for $(L,f)=(18.73\ \upmu\text{m},36.03\ \text{THz})$. \label{fig:deltagamma}}
\end{figure}

An important property that distinguishes the cavity from previous proposals is that the regions of high enhancement are not just the near fields around the cylinders. We now study the local enhancement as a function of position inside the cavity. We make use of the pointwise square norms of the two helicity components of the field $|\textbf{G}_+(\rr)|^2$, and $|\textbf{G}_-(\rr)|^2$, where $\sqrt{2}\textbf{G}_\pm(\rr)=\mathbf{E}(\rr)\pm \textrm{i} Z \mathbf{H}(\rr)$, with $Z$ being the medium impedance, are essentially the Riemann-Silberstein vectors \cite{Birula2013}. The ratio between the difference $|\textbf{G}_+(\rr)|^2-|\textbf{G}_-(\rr)|^2=\gamma(\rr)$ for the cases with [$\gamma_c(\rr)$] and without [$\gamma_0(\rr)$] the structure is the pointwise enhancement of the helicity density due to the cavity, which is identical to the ratio between the optical chirality density \cite{Tang2010} for those same cases. For an achiral system with properly chosen illumination the volume integral of $\gamma_c(\rr)/\gamma_0(\rr)$ is a good proxy for the CD enhancement \cite{Graf2019}. We assume that the cavity is filled with a medium characterized by $\epsilon_{\mathrm{r}}=2.14$, $\mu_{\mathrm{r}}=1$, and $\kappa=0$. The structure is illuminated with a left-handed polarized plane wave under perpendicular incidence. Its helicity intensities are $|\textbf{G}^0_+(\rr)|^2=C_+^0\neq 0$ and $|\textbf{G}^0_-(\rr)|^2=0$. Figure~\ref{fig:deltagamma} shows the spatial maps of $\gamma_c(\rr)/\gamma_0(\rr)=\gamma_c(\rr)/C_+^0$ for $(L,f)=(18.73\ \upmu\text{m},36.03\ \text{THz})$. We observe that the helicity density enhancement is large in electromagnetically large volumes, and that it increases when going away from the cylinders towards the center of the cavity, well outside the near-field regions around the cylinders.   

Finally, in \ifdefstring{\includeappendix}{no}{Sec.~\ref{app:acdachiral} of the Supplemental Material\footnote{See Supplemental Material [url], which includes Refs.~\cite{Stefanou2000,Mishchenko2016,Fruhnert2016b,Demesy2018,Garcia2018}.}}{}\ifdefstring{\includeappendix}{yes}{Appendix \ref{app:acdachiral}}{} we perform a quantitative analysis of the effects of violating the achirality requirement in our proposed cavity. We show that a nonzero ACD signal can be measured from an achiral analyte when the mirror symmetries of the cavity are broken due to a relative transverse displacement between the two arrays. This aspect needs to be considered in the manufacturing of such cavities.

In conclusion, we have presented a planar cavity that enhances the circular dichroism signal of chiral molecules by two orders of magnitude. The design exploits helicity-preserving cavity modes with long lifetimes that are mediated by the TE/TM reflection degeneracy in the limit of grazing incidence angle.
\begin{acknowledgments}
	This research has been funded by the Deutsche Forschungsgemeinschaft (DFG, German Research Foundation) under Germany's Excellence Strategy via the Excellence Cluster 3D Matter Made to Order (EXC-2082/1 -- 390761711), by the Helmholtz program ``Science and Technology of Nanosystems'' (STN), and by the European Union's Horizon 2020 research and innovation program under the Marie Sklodowska-Curie grant agreement No 675745. Finally, we are grateful to the company JCMwave for their free provision of the FEM Maxwell solver JCMsuite. 
\end{acknowledgments}

\ifdefstring{\includeappendix}{no}{\end{document}}{}
\ifdefstring{\includeappendix}{yes}{\appendix\setcounter{secnumdepth}{2}\renewcommand\thefigure{S\arabic{figure}}\section{Fresnel reflection coefficients at grazing incidence angles\label{app:fresnel}}
The Fresnel reflection coefficients for the TE and TM polarizations, found e.g. in \cite[Sec. 7.13]{Jackson1998}, can be written as
\begin{equation}
	\label{eq:tetm}
		\rte=\frac{1-\frac{Z_1}{Z_2}f\left(\beta,\frac{n_1}{n_2}\right)}{1+\frac{Z_1}{Z_2}f\left(\beta,\frac{n_1}{n_2}\right)},\ \rtm=\frac{1-\frac{Z_2}{Z_1}f\left(\beta,\frac{n_1}{n_2}\right)}{1+\frac{Z_2}{Z_1}f\left(\beta,\frac{n_1}{n_2}\right)},
\end{equation}
where $\beta$ is the angle of incidence measured from the normal, $(Z_{1,2},n_{1,2})$ are the impedances and refractive indexes of the two media, and $	f\left(\beta,\frac{n_1}{n_2}\right)=\sqrt{\frac{1}{\cos ^2\beta}-\left(\frac{n_1}{n_2}\right)^2\tan ^2\beta}$. It immediately follows from Eq.~(\ref{eq:tetm}) that, at grazing incidence when $\beta\rightarrow \pi/2$, then $\rte\rightarrow -1$ and $\rtm\rightarrow -1$, independently of  $(Z_{1,2},n_{1,2})$.
\section{Analysis of the double array cavity\label{app:array}}
Let us first consider a single array embedded in a non-absorbing achiral dielectric medium with $\epsilon_{\mathrm{r}}=2.14$ and $\mu_{\mathrm{r}}=1$. In a 2D hexagonal lattice, the real space(direct) hexagonal lattice vectors can be written as $\vec{a}_1=a\sqrt{4/3}[0,1]$, $\vec{a}_2=a\sqrt{4/3}[-\sin(\pi/3),\cos(\pi/3)]$. The corresponding momentum space(reciprocal) vectors are $\vec{b}_1=(2\pi/a)[\cos(\pi/3),\sin(\pi/3)]$, and $\vec{b}_2=(2\pi/a)[-1,0]$. As in any periodic 2D structure under plane wave illumination, the propagating and evanescent orders that can be excited are determined by the plane wave momentum components transverse to the array plane, and the discrete translational symmetry of the structure. For perpendicular illumination with transverse momentum $\vec{p}_{\perp}=[p_x,p_y]=[0,0]$, only orders with transverse momentum $\vec{p}_{\perp}^{\ nm}=[0,0]+n\vec{b}_1+m\vec{b}_2$ for any pair of integers $(n,m)$ can be excited. When the transverse momentum square $|\vec{p}_{\perp}^{\ nm}|^2$ is smaller than $k^2$, the squared wavenumber in the medium, the $(n,m)$ order is propagating, otherwise, it is evanescent. The longitudinal ($z$) component of the momentum corresponding to an $(n,m)$ order is
\begin{equation}
	\label{eq:pz}
	\begin{split}
		p_z^{nm}&=\pm \sqrt{k^2-|\vec{p}_{\perp}^{\ nm}|^2}\\
		&=\pm \sqrt{k^2-\left(\frac{2\pi}{a}\right)^2\left[n^2+m^2+nm\right]},
	\end{split}
\end{equation}
which can be real or imaginary, and where the $+/-$ sign indicates transmitted/reflected orders. 

Let us now consider a double array system embedded in the same medium as before. According to Eq.~(\ref{eq:slab}) of the main text, the modal condition in the cavity formed by the double array can be written as 
\begin{equation}
	\label{eq:wpl}
2Lp_z^{nm}+2\Phi^{nm}_{TE/TM}=2\pi l_{TE/TM},
\end{equation}
which, after using Eq.~(\ref{eq:pz}) and some manipulation, establishes a relation between $(n,m)$, $L$, $a$, and the wavelength in the medium $\lambda=2\pi/k$:
\begin{equation}
	\label{eq:lines}
	\begin{split}
		&L\sqrt{1-\left(\frac{\lambda}{a}\right)^2\left[n^2+m^2+nm\right]}\\
		&=\frac{\lambda}{2}\left(l_{TE/TM}-\frac{\Phi^{nm}_{TE/TM}}{\pi}\right).
	\end{split}
\end{equation}
We now consider any of the six $(n,m)$ pairs corresponding to the edges of the momentum space hexagon closest to the origin, i.e. the first six (degenerate) diffraction orders: $(0,1)$, $(1,0)$, $(0,-1)$, $(-1,0)$, $(1,-1)$, and $(-1,1)$. We assume that the corresponding angle $\beta$ is large enough to set $\Phi^{nm}_{TE/TM}=\pi$. Equation~(\ref{eq:lines}) becomes
\begin{equation}
	\label{eq:one}
	L\sqrt{1-\left(\frac{\lambda}{a}\right)^2}=\frac{\lambda}{2}(l_1-1).
\end{equation}
We are also interested in the $(n,m)=(0,0)$ modes, often referred to as Fabry-Perot modes, which correspond to perpendicular incidence ($\beta=0$). In such case and since $\Phi^{nm}_{TE}=\Phi^{nm}_{TM}=0$ (see \cite[Sec. 7.13]{Jackson1998}), we obtain from Eq.~(\ref{eq:lines})
\begin{equation}
	\label{eq:zero}
	L=\frac{\lambda}{2}l_0.
\end{equation}

Let us now establish the conditions for composite modes to arise when Eq.~(\ref{eq:one}) and Eq.~(\ref{eq:zero}) are simultaneously satisfied for a given pair of integers $(l_0,l_1)$. As shown in \cite[Sec. 2.3c]{Chang-Hasnain2012}, two different kind of modes can couple when the differential phase that they accumulate in a half round-trip is an integer multiple of $2\pi$. Since the left hand side of Eq.~(\ref{eq:wpl}) is the phase accumulated in a round trip, we easily obtain the phase accumulated in a half round trip as
\begin{equation}
	\begin{split}
	&\text{Zeroth-order: }Lp_z^{00}=\pi l_0,\\
	&\text{First-order: }Lp_z^{10}+\pi=\pi l_1,\\
	\end{split}
\end{equation}
from where it follows that the two kinds of modes can couple when $l_1-l_0$ is even or, equivalently, when $(l_1-1)-l_0$ is odd.

Finally, we seek an expression for the angle $\beta^{nm}$ corresponding to the $(n,m)$ diffraction order. First, it is readily derived from Eq. (\ref{eq:wpl}) that the angle $\beta^{nm}$ can be calculated as
\begin{equation}
	\label{eq:beta1}
	\beta=\arccos\frac{p^{nm}_z}{k}=\arccos\left(\frac{\left(l-\Phi/\pi\right)c}{2f^{nm}L}\right),
\end{equation}
where $c$ is the speed of light in the medium. Then, for a given $(n,m)$ mode, the modal frequency $f^{nm}$ can be deduced from Eqs. (\ref{eq:pz}) and (\ref{eq:wpl}) to be
\begin{equation}
	\label{eq:fnm}
	f^{nm}=c\sqrt{\left(\frac{l-\Phi/\pi}{2L}\right)^2+\frac{n^2+m^2+nm}{a^2}}.
\end{equation}
We combine Eqs. (\ref{eq:beta1}) and  (\ref{eq:fnm}) to obtain
\begin{equation}
	\label{eq:beta}
	\begin{split}
		&\beta^{nm}=\\
		&\arccos\left(\frac{l-\Phi/\pi}{\sqrt{\left(l-\Phi/\pi\right)^2+\left(\frac{2L}{a}\right)^2\left[n^2+m^2+nm\right]}}\right).
	\end{split}
\end{equation}
\section{Transmittance CD\label{app:forwardonly}}
Results shown in Figs.~\ref{fig:cdehnforward}-\ref{fig:t0}.
\begin{figure}[h!]
		\includegraphics[width=\linewidth]{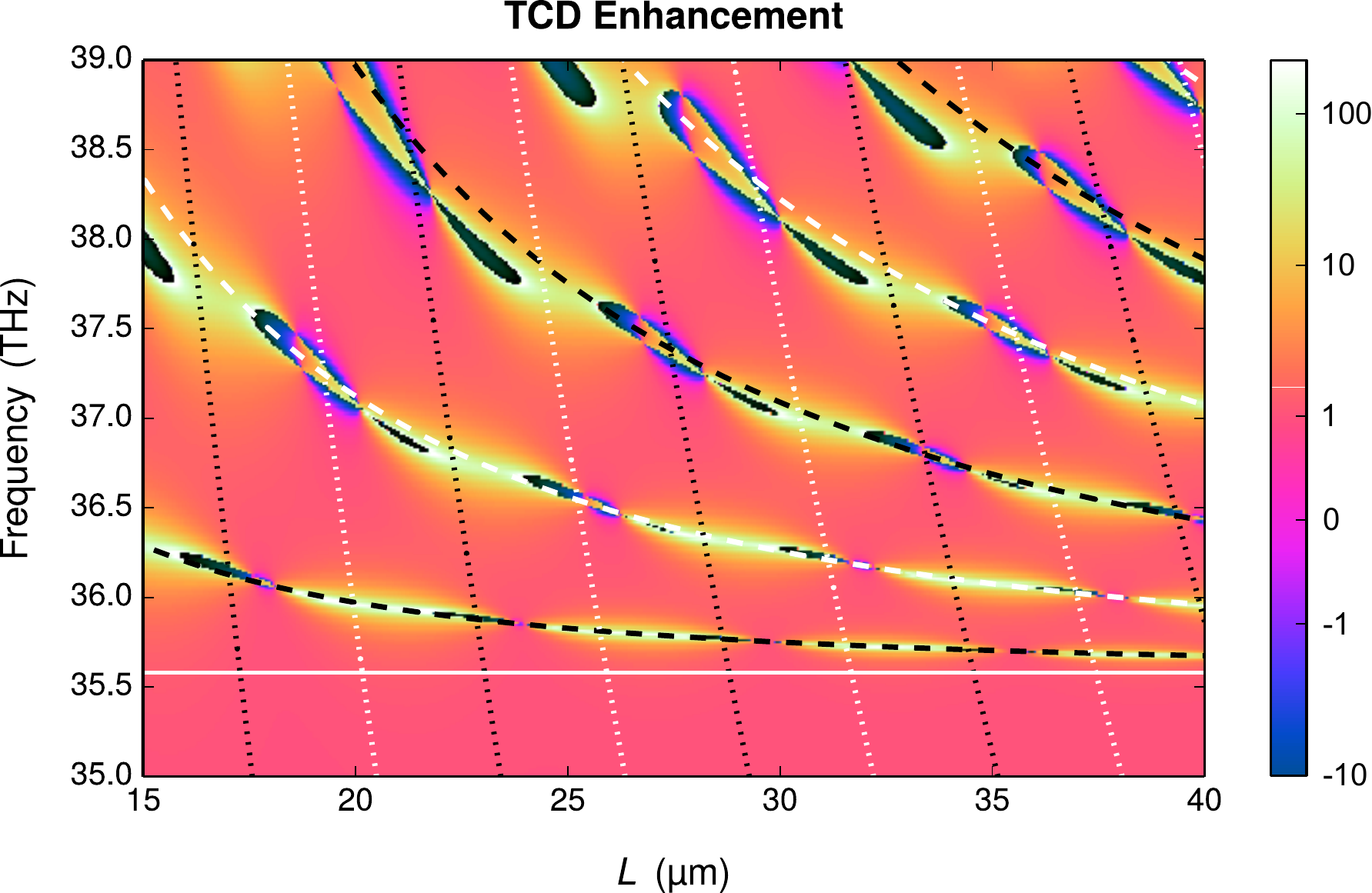}
	\caption{Transmittance CD enhancement for the same system and conditions that result in the ACD enhancements shown in Fig.~\ref{fig:lines}(b). The scale is a symmetric logarithmic scale: The logarithm of the displayed quantity for positive values larger than 1, the negative logarithm of the absolute value of the quantity for negative values smaller than -1 and a linear interval from -1 to 1. \label{fig:cdehnforward} }
\end{figure}
\begin{figure}[h!]
		\includegraphics[width=\linewidth]{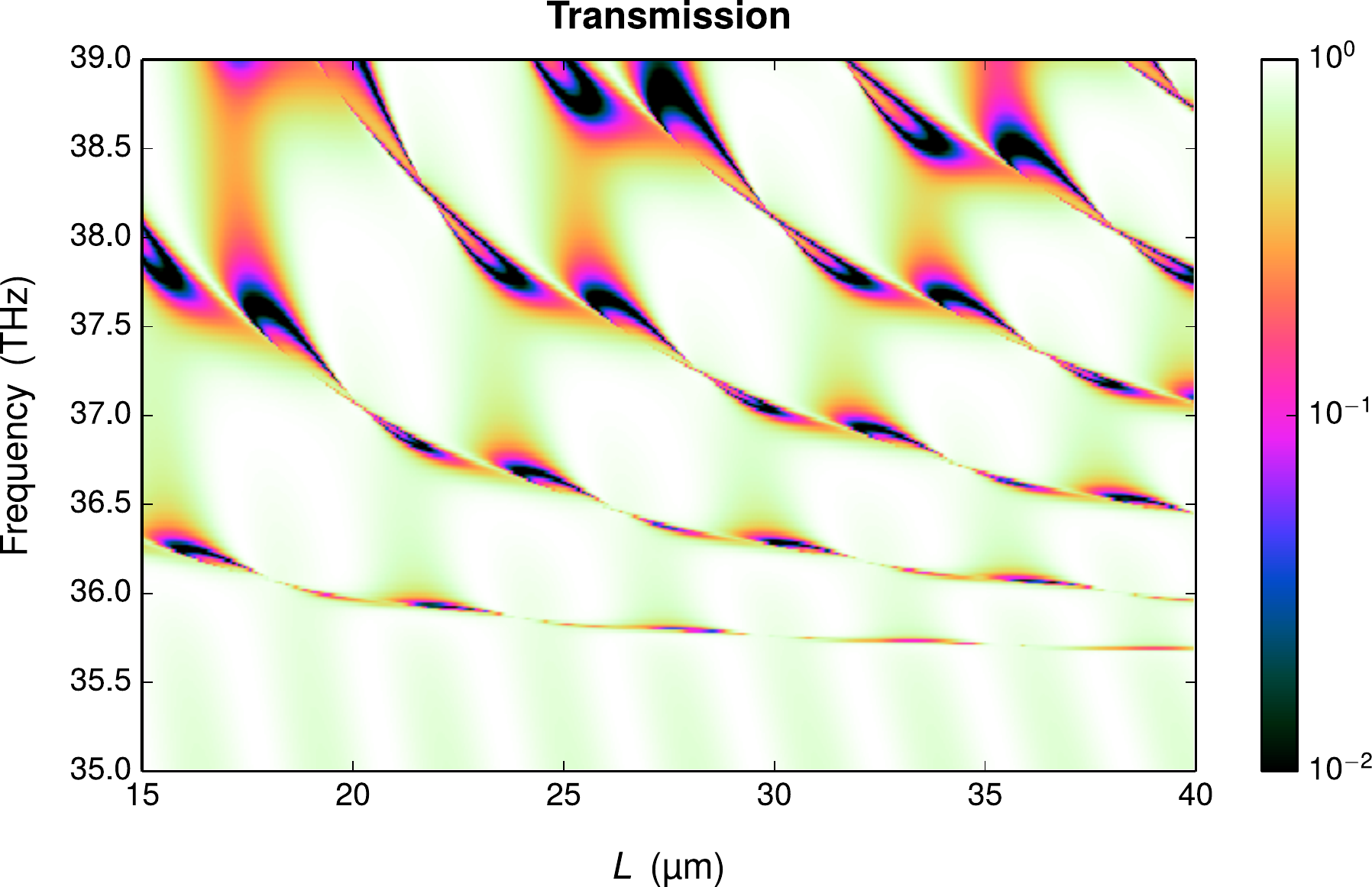}
	\caption{Average of the transmittance under LCP and RCP illumination $\left(P^{\text{fwd}}_++P^{\text{fwd}}_-\right)/2$. \label{fig:t0} }
\end{figure}
\section{Non-zero ACD signal from an achiral analyte due to breaking the achirality requirement\label{app:acdachiral}}
Let us assume that we want to use a structure for enhancing the CD signal of a molecular solution. In order to ensure that the CD signal depends only on the chiral absorption properties of the molecules and not on their non-enantiospecific achiral absorption properties the structure must be achiral, and the two subsequent illuminations used for the CD measurements must transform into each other by one of the spatial inversion symmetries of the structure \cite{Graf2019}. We now perform a quantitative analysis of the effects of violating the achirality requirement in our proposed cavity. The requirement is violated by introducing a relative transverse displacement between the two arrays. As result, a non-zero ACD signal can be measured from an achiral analyte. 
\begin{figure}[h!]
		\includegraphics[width=\linewidth]{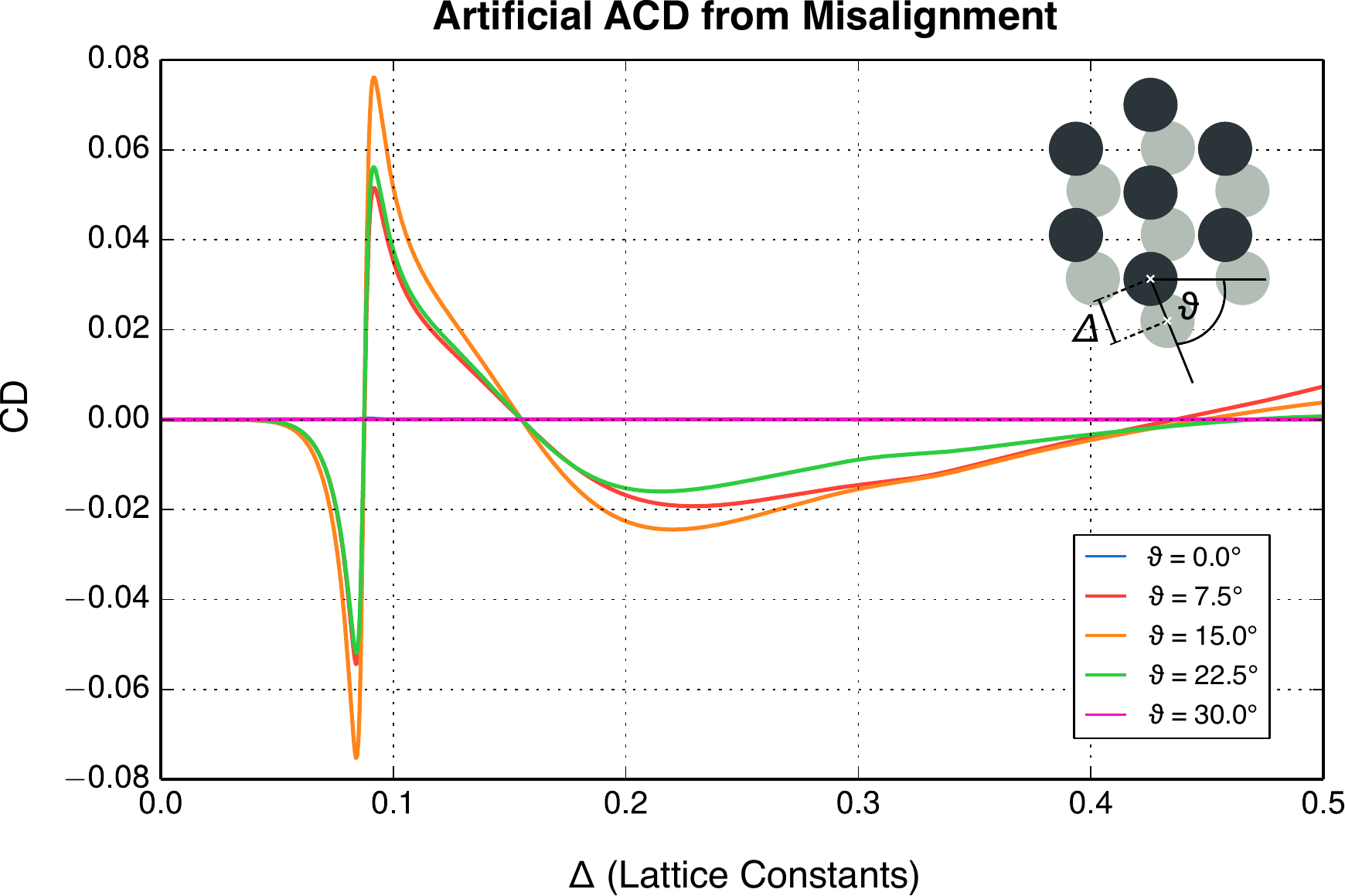}
	\caption{Absorbance CD from an achiral absorbing analyte due to the breaking of the achirality requirement by the double array cavity. A relative transverse displacement between the two arrays ($\Delta$) breaks the necessary mirror symmetries for most displacements. The different lines correspond to different displacement directions, as shown in the inset.\label{fig:mbreakings} }
\end{figure}
The medium in the cavity is now absorptive and achiral, with the same parameters as in Eq.~(\ref{eq:consrel}) of the main text except that $\kappa$ is now set to zero. The length of the cavity $L$ and frequency $f$ are $(L,f)=(18.73\ \upmu\text{m},36.03\ \text{THz})$. Figure~\ref{fig:mbreakings} shows the ACD signal computed with Eq.~(\ref{eq:relcd}) of the main text as a function of $\Delta$: The transverse displacement in units of the lattice constant. The different lines correspond to displacements along different directions defined by their angle $\theta$ with respect to the horizontal axis [see inset]. For $\theta=\{0,30\}$ degrees the ACD signal is zero for all displacements because the structure still has a mirror symmetry through a plane containing the optical axis, which transforms a perpendicularly incident left-handed plane wave into a perpendicularly incident right-handed plane wave. For $\theta=\{7.5,15,22.5\}$ degrees the ACD signal is non-zero for most non-zero displacements, and three sign changes (zero-crossings) can be observed. For each displacement direction, the zero-crossing that occurs near $\Delta=0.45$ coincides with the point for which the displacement results in the recovery of a geometrical mirror symmetry. The point is slightly different for each $\theta$. A sharp feature is observed at $\Delta \approx 0.087$. The corresponding TCD results are similar. We remark that the only absorption in the system is that of the analyte.

These results highlight the importance of meeting the achirality condition in structure-illumination systems for enhanced CD measurements. In practical applications, the unavoidable deviation from this condition will set a limit on the achievable sensitivity. This notwithstanding, the effect can be also be useful, for example, to correct small misalignments between the two arrays by minimizing the CD signal of an achiral absorbing medium like ethanol.   

\section{Numerical simulations\label{app:code}}
The numerical results contained in this Letter have been produced using a code developed in house. The code rigorously solves the electromagnetic interaction between any incoming plane wave illumination and planar multilayer systems extending to infinity in two orthogonal directions. Each of the different layers in a given system can be either an isotropic and homogeneous semi-infinite medium or slab, or a two dimensional array of identical scatterers. In the former cases, the solution makes use of the Fresnel transmission and reflection formulas. In the latter case, the self-consistent problem of a plane wave illuminating a 2D array of electromagnetically coupled particles is solved by means of multiple scattering techniques \cite{Stefanou2000}. Upon reaching a given layer, the external illumination will excite a set of forward and backward propagating and evanescent plane wave orders, which will at their turn contribute to the excitation of the other layers. Finally, the self-consistent solution considering the external illumination and the interaction between all the layers is obtained taking into account all the orders below a chosen cutoff point. The final outcome of the calculations are the complex weights of the plane waves existing in each of the layers, including the first and last ones which represent the two semi-infinite media that sandwich the multilayer system.

We now provide more details about the methods, explain the sources of error in the calculations, and show the result of sanity checks.

While the isotropic and homogeneous media layers are trivially solved, two dimensional arrays of particles are more challenging. The main ingredient for their solution is the transfer matrix operator, a.k.a T-matrix, of an isolated individual particle. The T-matrix is a very common tool in physics and engineering \cite{Mishchenko2016} which allows to compute the electromagnetic response of an object under any illumination. Once the T-matrix of an individual particle is known, e.g. for a single silicon cylinder in this Letter, the translation theorems of multipolar fields are used in solving for the total response, which includes the interaction of the array with the externally incident plane wave and the multiscattering events between the particles. The publicly available code described in Ref. \onlinecite{Stefanou2000} can only handle arrays of spherical particles. Our code allows for arrays of particles of any shape. The main difference is that, while the T-matrix of a sphere can be obtained analytically from its Mie coefficients, the T-matrix of a general particle needs to be computed numerically. In our case, we use an approach similar to the one described in Ref. \onlinecite{Fruhnert2016b}, where the T-matrix of an isolated object is obtained by exciting it with several different incident fields and collecting the scattered field for each instance. Instead of the plane wave excitation used in Ref. \onlinecite{Fruhnert2016b}, we here use multipolar fields as in Ref. \onlinecite{Demesy2018}. The numerical calculations of the T-matrix of the silicon cylinders have been performed with the commercially available JCMSuite software, which is a finite element method solver (FEM) of Maxwell's equations. We use their built-in implementation of Ref. \onlinecite{Garcia2018} to perform the decomposition of the scattered field into vector spherical wave functions. The simulations are performed using a 2D layout with cylindrically symmetric boundary conditions, allowing a faithful geometrical representation of the cylinder while requiring much lower computational requirements than its 3D counterpart. 

The sources of error in the calculations are the following ones. The accuracy of the FEM solution, the need to select a maximum multipolar order for the T-matrix of the individual cylinders, and the cutoff of the number of evanescent plane wave orders that are considered in the self-consistent calculations between layers. We have selected the corresponding control parameters so that the results converge to a stable solution which does not vary when the control parameters are changed to further improve the numerical precision. In particular, we have selected an FEM mesh size equal to 1/5-th of the wavelength inside the silicon and polynomial degree equal to 6, we compute the individual T-matrices up to multipolar order 4, and all the plane wave orders whose transverse momentum is smaller or equal than three times the wavenumber are included in the multilayer calculations.  

As a sanity check, we have used COMSOL to compute the ACD enhancement at three points in Fig.~\ref{fig:lines}(b). Table \ref{tab:comp} shows the results. The match is very satisfactory, with a maximum relative discrepancy of $\approx$1\% at the resonant point.
\begin{table}[h!]
\begin{center}\begin{tabular}{cccc}
		$L$ ($\upmu$m)& $f$ (THz)  & Enhancement & Enhancement (COMSOL) \\
		\hline
		18.722 & 36.027 & 125.822& 124.575\\
		18.722 & 35.397&  0.875& 0.876\\
		24.600 & 39.000 &  -4.067& -4.067\\
\end{tabular}\end{center}
		\caption{
		Absorbance CD enhancement at three points in the cavity length ($L$) and frequency ($f$) space computed with our own code and COMSOL.\label{tab:comp}}
\end{table}
We note that the speed improvement factor of our code with respect to COMSOL or JCM is around 100, which allows us to perform the kind of study presented in this Letter in reasonable time.
\section{Reflection off a single array\label{app:figures}}
Results shown in Fig.~\ref{fig:reflectionsinglearray}.
\subsection{Average number of reflections in a mode\label{app:bounce}}
The evolution of the energy contained in the mode of a planar cavity can be seen as successive reflections on the top and bottom walls. At each reflection event, a fraction of the energy reflects into the same mode and the rest ``leaks out'' through other channels. We denote by $p$ the fraction of the total energy that stays in the same mode. For the double array cavity in the main text, $p=|\hcons|^2+|\hflip|^2$. The quality factor of the mode or, equivalently, its lifetime, must be proportional to the average number of reflections before the energy leaves the mode. We can compute such average by taking $p$ as the probability of staying in the mode at each reflection, and considering that whether the energy leaves the mode or not at a given reflection event is independent of what happened in previous events. Then, the probability of bouncing $n$ times before leaving is $p^{n}(1-p)$. The sought after average is then  
\begin{equation}
	N=\sum_{n=1}^{\infty} np^{n}(1-p)=(1-p)\sum_{n=1}^{\infty} np^{n},
\end{equation}
which we manipulate into
\begin{equation}
	N=(1-p)p\sum_{n=1}^{\infty} np^{n-1}=(1-p)p\sum_{n=1}^{\infty}\frac{d}{dp}p^{n},
\end{equation}
so that we can interchange the derivative and the sum, solve the sum explicitly, take the derivative, and reach the final result:
\begin{equation}
	N=(1-p)p\frac{d}{dp}\frac{p}{1-p}=(1-p)p\frac{1}{(1-p)^2}=\frac{p}{1-p}.
\end{equation}
\onecolumngrid
\begin{figure*}
		\includegraphics[width=0.90\textwidth]{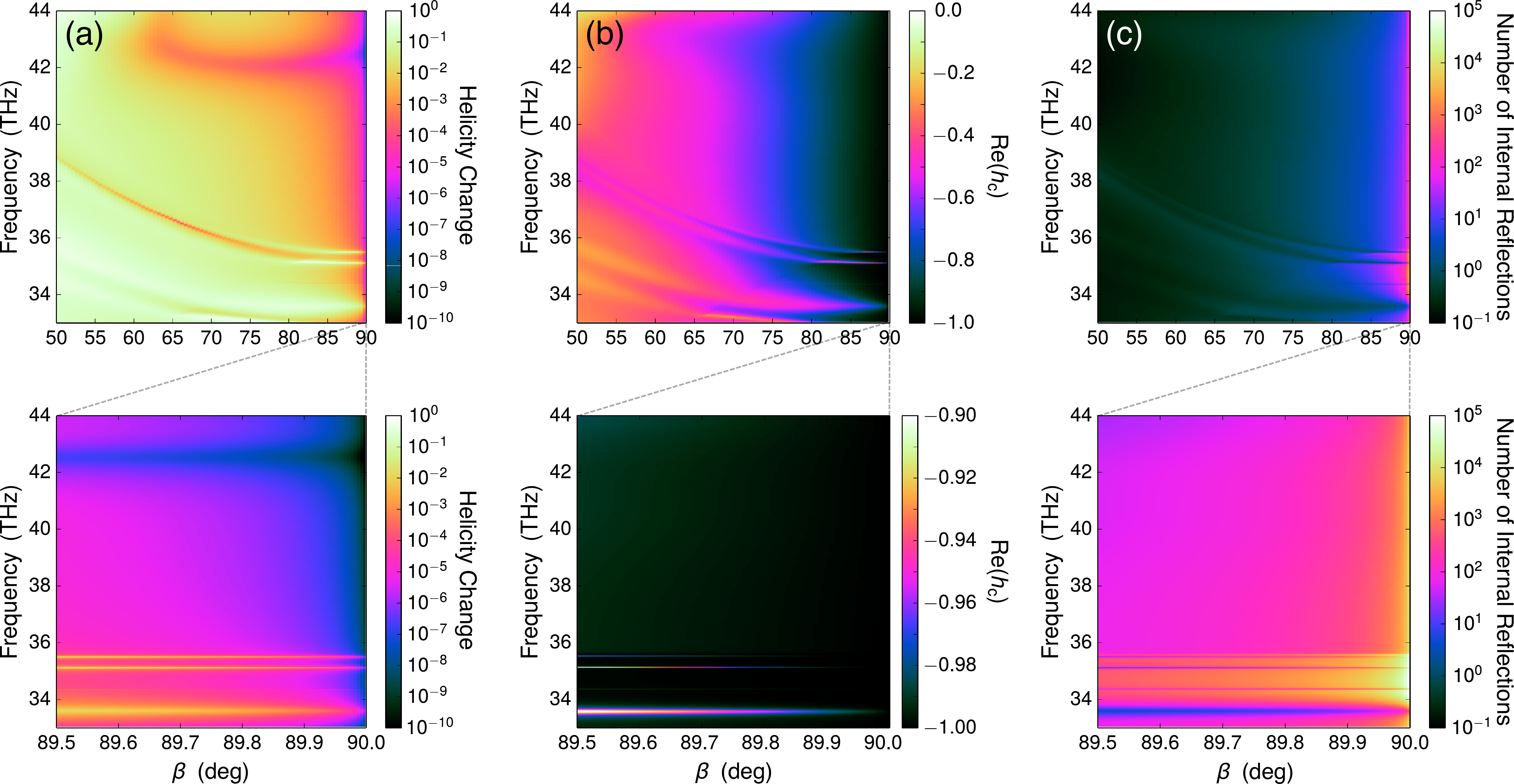}
		\caption{Frequency and angle of incidence ($\beta$) analysis of the zeroth-order reflection off a single array like the ones in Fig.~\ref{fig:lines}(a), immersed in an $\epsilon_{\text{r}}=2.14$ medium. (a) Helicity change measure $|\hflip|^2/\left(|\hcons|^2+|\hflip|^2\right)$. (b) Real part of $\hcons$. (c) Average number of internal reflections before the energy leaves the mode (see text).\label{fig:reflectionsinglearray} }
\end{figure*}
 \end{document}}{}
\end{document}